\documentclass[sigconf]{acmart}

\usepackage{booktabs} 

\usepackage{amsmath,amssymb}
\usepackage{float,color}
\usepackage{textcomp, libertine}

\newcommand{\pbid}[1]{p_{\text{bid},#1}}
\newcommand{\nshares}[1]{N_{\text{shares},#1}}
\newcommand{\volpref}[1]{\nu_{#1}}
\newcommand{\mech}[1]{\mathcal{M}_{\text{#1}}}

\setcopyright{acmcopyright}

\acmDOI{10.1145/3321707.3321734} 
\acmISBN{978-1-4503-6111-8/19/07} 
\acmConference[GECCO '19]{the Genetic and Evolutionary Computation Conference 2019}{July 13--17, 2019}{Prague, Czech Republic}
\acmYear{2019}
\copyrightyear{2019}

\acmPrice{15.00}

\begin{document}
\title{Selection mechanisms affect volatility in evolving markets}

\author{David Rushing Dewhurst}
\authornote{To whom correspondence should be addressed.}
\orcid{0000-0001-6130-1833}
\affiliation{%
  \institution{University of Vermont}
  \streetaddress{210 Colchester Avenue}
  \city{Burlington} 
  \state{Vermont} 
  \postcode{05405}
}
\email{david.dewhurst@uvm.edu}

\author{Michael Vincent Arnold}
\affiliation{%
  \institution{University of Vermont}
  \streetaddress{210 Colchester Avenue}
  \city{Burlington} 
  \state{Vermont} 
  \postcode{05405}
}
\email{michael.arnold@uvm.edu}

\author{Colin Michael Van Oort}
\affiliation{%
  \institution{University of Vermont}
  \streetaddress{210 Colchester Avenue}
  \city{Burlington} 
  \state{Vermont} 
  \postcode{05405}
}
\email{cvanoort@uvm.edu}

\renewcommand{\shortauthors}{D.R. Dewhurst et al.}

\begin{abstract}
Financial asset markets are sociotechnical systems 
whose constituent agents are subject to evolutionary 
pressure as unprofitable agents exit the marketplace and
more profitable agents continue to trade assets.
Using a population of evolving zero-intelligence agents and a 
frequent batch auction price-discovery mechanism as substrate, we 
analyze the role played by evolutionary selection mechanisms
in determining macro-observable
market statistics. 
Specifically, we show that selection mechanisms incorporating a 
local
fitness-proportionate component are associated with 
high correlation between a 
micro, risk-aversion parameter and a commonly-used 
macro-volatility statistic, while a purely quantile-based 
selection mechanism shows significantly less correlation 
and is associated with higher absolute levels of fitness (profit) than 
other selection mechanisms.
These results point the way to a possible restructuring of market incentives 
toward reduction in market-wide worst performance,
leading profit-driven agents to behave in ways that are associated with
beneficial macro-level outcomes.
\end{abstract}

%
%



\maketitle

\section{Introduction}
\begin{figure*}[!htp]
    \centering
    \includegraphics[width=0.7\textwidth]{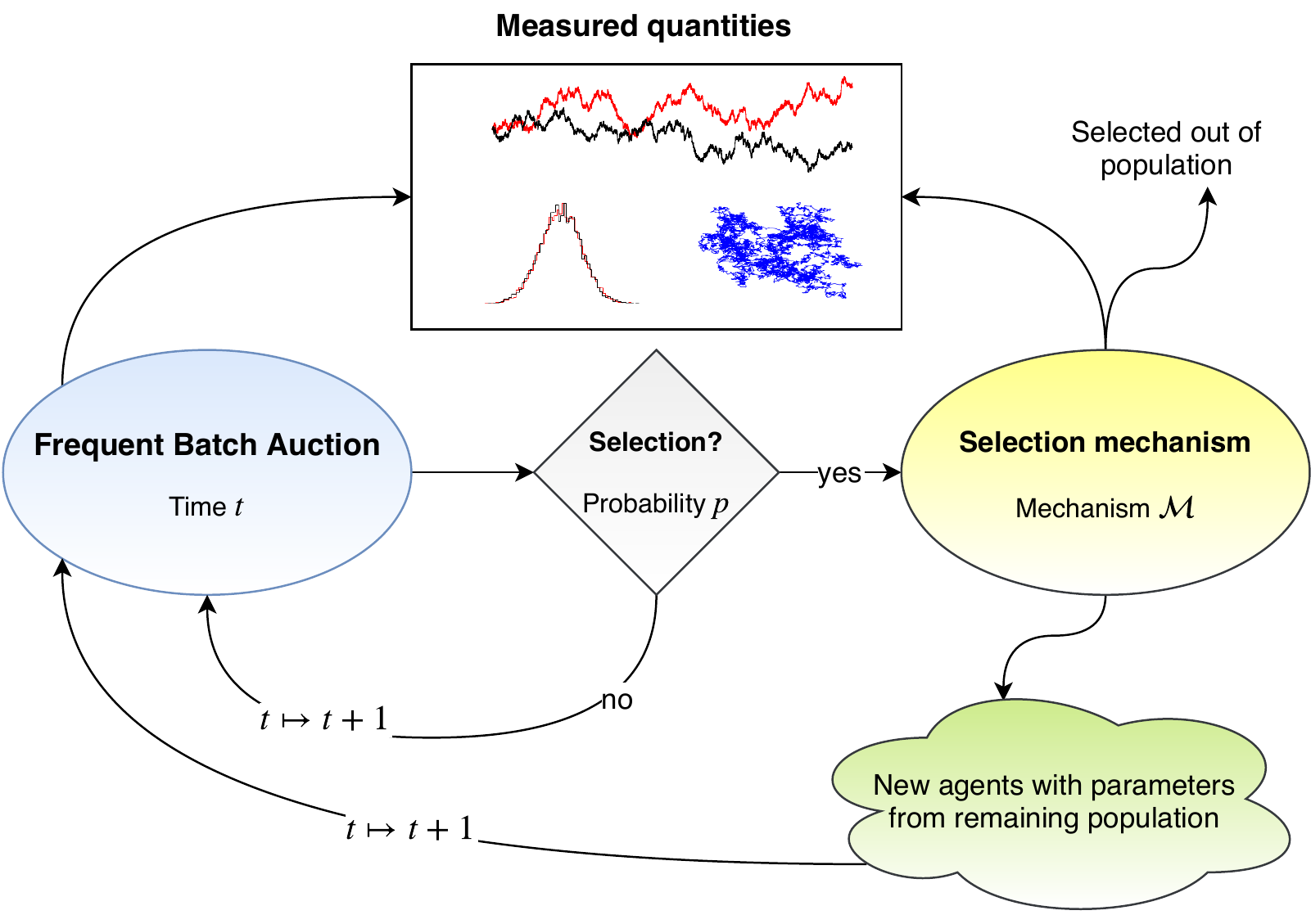}
    \caption{A cartoon of the financial system considered here is shown.
    Agents interact via the mechanism of a frequent batch auction, 
    explained in Section \ref{sec:price-discovery},
    and are subject to a type of probabilistic selection mechanism
    that discards agents with low fitness, which is here defined by profit,
    and replaces discarded agents with new agents whose parameters are drawn from the distribution of parameters among remaining agents.
    Statistics from market activity and the selection process are gathered 
    during iterations of the simulation and subsequently analyzed.}
    \label{fig:explanation}
\end{figure*}
The concept of adaptive financial markets has been studied 
extensively in quantitative finance for nearly twenty years.
The efficient markets hypothesis (EMH), which in its weakest form states that 
the price of an asset should, under conditions including costless information and agents
with rational expectations about the future,
reflect all publicly-available past information, has been an influential starting point for 
the study of financial theory since its initial publication in the late 1960s 
\cite{malkiel1970efficient}.
However, there is empirical evidence that this hypothesis does not hold.
A well-documented momentum effect exists for asset prices: assets that have done 
well (poorly) in past time periods will tend to do well (poorly) in future time periods, for 
periods ranging up to a year in the future \cite{jegadeesh1993returns}.
In addition, there have been objections to the rational expectations assumption of EMH on a theoretical basis
\cite{lo2004adaptive,lo2005reconciling,kim2011stock}.
Critics of the EMH have proposed a so-called ``adaptive-markets hypothesis'' (AMH), in the framework 
of which the population of agents is in constant flux, adapting to changing market 
forces and subject to evolutionary pressure \cite{farmer1999frontiers}.
The rise of high-frequency trading (HFT) in response to a shift in the
regulatory environment in U.S.\ asset markets in the mid-2000s is one factor that has lent credence to the AMH theory 
\cite{johnson2012financial,smith2010high,menkveld2013high}.
\medskip

\noindent
\begin{figure*}[!h]
\centering
	\includegraphics[width=\textwidth]{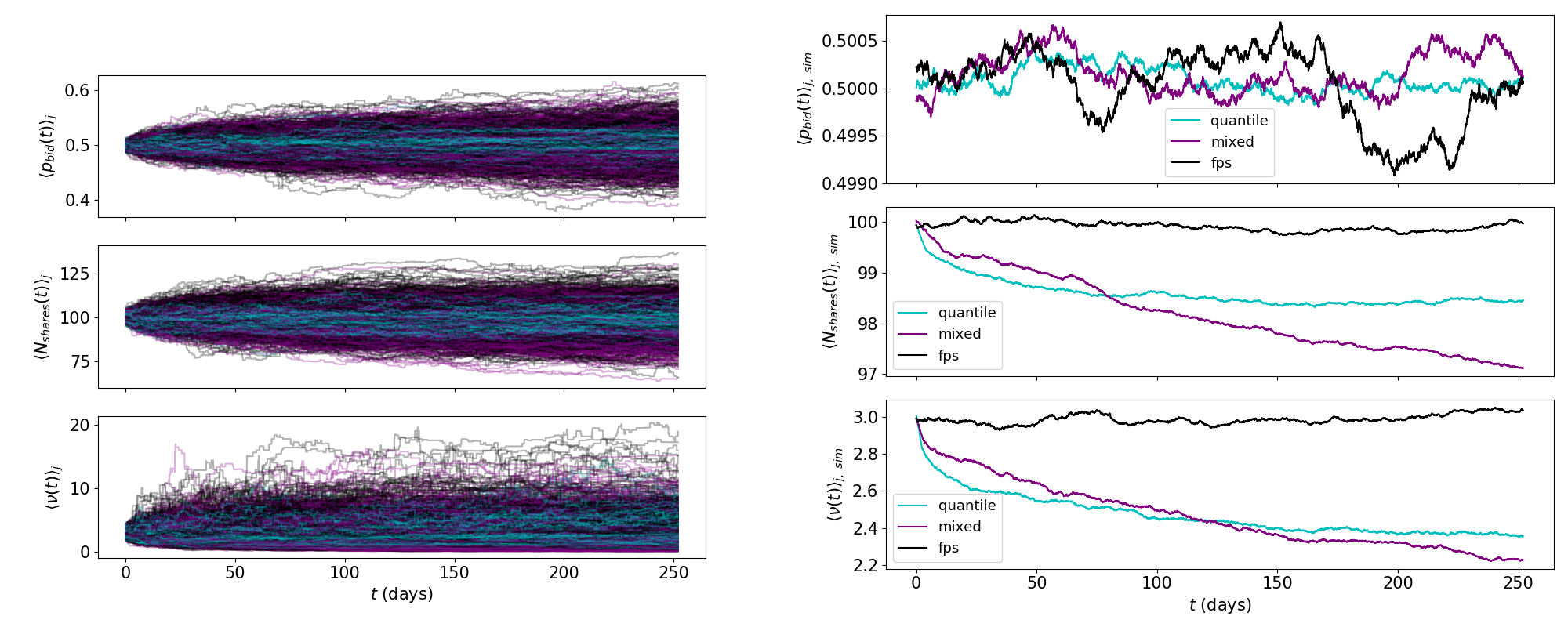}
	\caption{Means and standard deviation of parameter time series differ by selection 
	mechanism.
	The left panel displays parameter time series averaged over agents; a single 
	time series is plotted for each run of the simulation.
	The right panel displays parameter time series averaged over both agents and 
	runs of the simulation.
	Overall, the quantile selection mechanism leads to lower spatial standard deviations 
	across runs of the simulation, 
	as can be observed in the left panel.
	While both the quantile and mixed selection mechanisms show decaying average $N_{\text{shares}}$
	and $\nu$, fitness-proportionate selection shows no such behavior.
	The fitness-proportionate selection mechanism shows larger variation across runs of the simulation 
	in these variables as well, with much larger extreme values of $\nu$ than either of the other 
	mechanisms.
	When averaged over both agents and runs of the simulation, $p_{\text{bid}}$ shows effectively 
	no variation in time.}
	\label{fig:params-tile}
\end{figure*}
As a result of the apparent adaptive nature of modern financial markets, 
there has been substantial application of agent-based model (ABM) methods
to model various market features of interest \cite{lebaron2000agent,savit1999adaptive,
hommes2002modeling}.
Such models often assume constant a particular selection mechanism by which agents of low fitness 
(usually, low profitability) are selected out of the market and agents of higher fitness remain 
\cite{zhang1998modeling,kinoshita2013evolutionary,hommes2009complex}.
However, the design of the selection mechanism may have a material effect on measurable quantities in the 
marketplace, such as price or return time series, preferences (parameters) of high-fitness agents, 
and volatility.
\medskip

\noindent
In this work, we analyze the role of various selection mechanisms in determining the preferences of a population 
of evolving zero-intelligence agents interacting through the means of an auction mechanism.
Comparing two fundamentally distinct mechanisms---one a global mechanism based on population profit quantiles and the other
a local mechanism based 
on sample profitability---we show that this choice not only affects the dynamic behavior and 
distribution of agent parameters 
as shown in Figures \ref{fig:params-tile} and \ref{fig:params-dists}, but also has a significant effect on 
micro-macro volatility correlations. 
We find that incorporating local fitness-proportionate selection greatly increases the correlation between 
a micro-level, risk aversion parameter and macro-level volatility as measured by standard financial econometric 
machinery, compared to purely quantile-based selection.

\section{Theory and simulation}
We focus our attention on the mechanism by which agents of low fitness---unprofitable agents---
are selected out of the market.
In real-world financial markets, agents whose trading strategies produce 
low returns on capital can experience an outflow of funds to agents whose strategies 
produce better returns as investors seek the highest possible return subject to their risk preferences.
In a world of perfect information, firms would thus be selected out of a market according to a type of 
fitness-proportionate selection.
Real financial markets---and markets of all kinds---are rife with 
information asymmetries \cite{akerlof1978market,kim1991market}; here, we focus on the situation 
of perfect information to highlight the importance of the selection mechanism on macro-level observables.
\subsection{Agent specification}
Agent $i$'s fitness function at time $t$ is given by its profit at that time, defined as 
\begin{equation}
	\pi_i(t) = c_i(t) + s_i(t)X(t),
\end{equation}
where $c_i$ and $s_i$ are the amount of cash held by agent $i$ (units of currency),
and number of shares of the asset held by agent $i$, respectively,
and $X$ is the price of the asset.
Agents are permitted to ``sell short": they are not restricted to have a non-negative amount of cash.
Agents are zero-intelligence \cite{gode1993allocative,farmer2005predictive}
in the sense that their actions are purely random given a set of parameters; 
agents do not adapt in our model but are subject to evolutionary pressure across generations.
The behavior of an agent is determined by three parameters:
$\pbid{i}$, the probability of submitting a bid order in a time period given that the agent trades
in that time period; $\nshares{i}$, the mean number of shares submitted by the agent in a time period;
and $\volpref{i}$, the so-called ``volatility preference" of the agent, the role of which we will describe 
presently.
Given the asset price at time $t$, $X(t)$, the agent submits a bid order with probability 
$\pbid{i}$ (equivalently, an ask order with probability $1 - \pbid{i}$) 
with number of shares distributed as $N_i(t) \sim \text{Poisson}(\nshares{i})$ 
and price distributed according to the random variable 
\begin{equation}\label{eq:agent-order}
	X^{(\text{order})}_i(t+1) = X(t) + \volpref{i}u_i(t),
\end{equation}
where $u_i(t) \sim \mathcal{U}[-1,1]$.
The volatility preference parameter thus encodes a measure of regard for the current price level 
$X(t)$: low $\volpref{i}$ implies a preference for the current price level, while larger values 
lead to larger moves in both positive and negative directions.
This parameter is interpreted as a measure of risk aversion (small $\nu$) or risk neutrality / risk 
seeking (large $\nu$).
 \begin{figure*}[!htp]
\centering
	 \includegraphics[width=\textwidth]{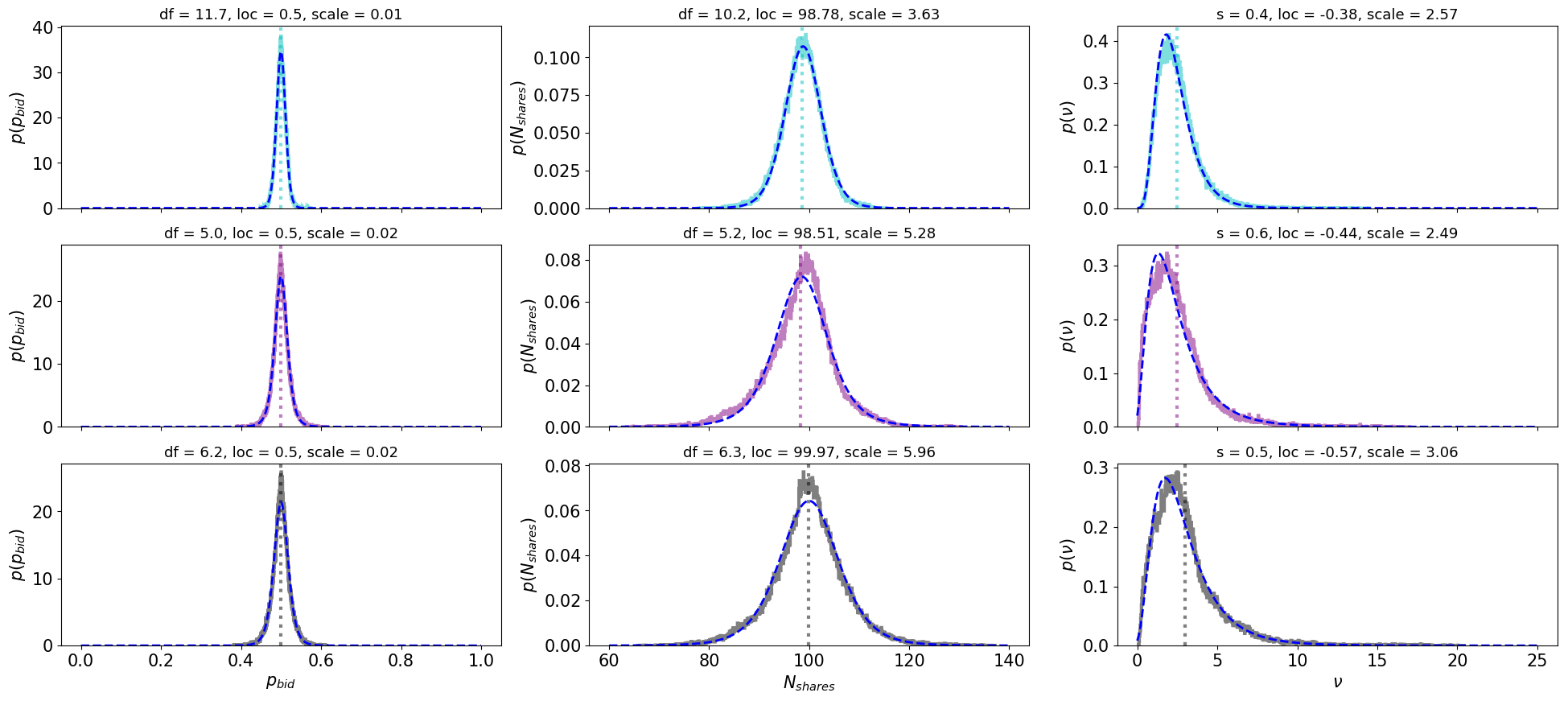}
	 \caption{When uncoupled from time, distributions of parameters are similar across 
	 selection mechanisms.
	 These distributions are calculated by computing the empirical pdfs over the union of time 
	 series of parameters over all 
	 points in time and runs of the simulation. 
	 The mixed selection mechanism displays the heaviest tails in the distributions of 
	 $p_{\text{bid}}$ and $N_{\text{shares}}$, followed by the fitness-proportionate mechanism.
	 From top to bottom: the quantile-based, mixed, and fitness-proportionate mechanisms.
	 The blue dashed curves and titles indicate optimal fits to the empirical distributions 
	 as computed using maximum likelihood estimation.
	 The distributions of $p_{\text{bid}}$ and $N_{\text{shares}}$ are well-fit by a ${\bf t}$-distribution,
	 while the distribution of $\nu$ is well-fit by a log-normal distribution.
	 }
	 \label{fig:params-dists}
 \end{figure*}
\subsection{Price-discovery mechanism}\label{sec:price-discovery}
Market price is determined by a frequent batch auction (FBA), introduced by Budish {\it et al.}\ 
as a response to HFT strategies \cite{budish2015high,budish2014implementation},
which we now describe briefly.
Modern financial markets primarily use a continuous double auction
(CDA) mechanism to match buyers and sellers, 
though FBA has recently attracted much theoretical and intellectual property interest
\cite{cushing2013automated,wah2016strategic}, 
and batch auctions more generally have been in use since at least 2001 on the Paris 
Bourse \cite{muscarella2001market}. 
CDA and FBA share several attributes. Both mechanisms
are double-sided mechanisms in which any number of buyers and sellers may participate, and participants may enter or leave the market at any time under both mechanisms.
Both mechanisms also maintain an order book, which accumulates orders that have not yet been executed.
In practice, 
both mechanisms feature a similar price-time execution priority for resting orders, 
though the implementation may vary slightly.
In other words, orders that have a better price, bids with higher prices or asks with lower prices, are executed first.
Ties in price are broken by the age of the order, with older orders executing first.
\medskip

\noindent
CDAs allow agents to submit orders at any time, 
and these orders are immediately matched against resting orders if possible.
Orders that are not immediately executed will be added to the order book,
where they will wait for a counter-party to accept their conditions.
This procedure results in trading that occurs continuously, aligning with the name of the mechanism.
On the other hand, FBAs divide trading into discrete intervals.
Within each interval agents may submit orders at any time, which are then placed in the order book.
At the end of a trading interval,
a single uniform execution price is selected by locating the intersection of the supply and demand curves 
(i.e. price and quantity of orders from both sides of the market are used to identify the execution price).
Orders to buy with a limit price at least as high as the selected execution price 
and orders to sell with a limit price at least as low as the selected execution price 
are then eligible to execute.
Eligible orders are then matched together following price-time priority, i.e. bids with higher prices and asks with lower prices are matched first, with ties broken by order age, 
and further ties broken by uniform random selection.
The orders that did not execute at time $t$ remain in the book and are reconsidered for execution in future
time periods until such time as the matching engine considers them to be ``stale", or too old for 
consideration. 
The implementation of FBA considered here sets the maximum allowed time for an order to remain in the book 
to be 24 time periods, or one day.
\medskip

\noindent
Since the aim of this work is to understand the effects of selection pressure and different selection 
mechanisms on macro-statistics of market activity, we attempt to abstract away other details of 
real-world asset markets. 
Though the U.S. National Market System (NMS) is a fragmented market with no fewer than thirteen exchanges
 operating at time of writing \cite{o2011market},
 we consider only a single exchange and matching engine here. 
 As noted above, agents are effectively zero-intelligence; though they are subject to selective pressure 
 and thus the population of agents may become more profitable over time as weak agents are selected out,
 individual agents do not adapt to changing market circumstances.
\subsection{Selection mechanisms}
 Selection occurs with constant probability of $p_{\text{selection}} = \frac{1}{24}$ each time period, so that
 there is a selection event in one out of every 24 time periods (hours) on average.
 We consider three selection mechanisms: a quantile-based mechanism (truncation selection),
 denoted by $\mech{quantile}$; a type of fitness-proportionate selection, $\mech{fps}$,
 and a mixture of the two mechanisms, $\mech{mixed}$, each of which is a well-known selection method
 \cite{blickle1996comparison}.
 The quantile-based mechanism removes agents $i$ whose profit satisfies 
 $\pi_i(t) < F^{\leftarrow}_{\pi(t)}(q)$, where $q$ is a quantile (number between 0 and 1) and 
 $F^{\leftarrow}_{\pi(t)}$ is the quantile function of the profit distribution across all agents active at 
 time $t$.
 We set $q = 0.1$ to remove the bottom 10\% of agents each time the quantile-based mechanism is activated.
 The fitness-proportional selection mechanism is a standard implementation of such a procedure:
 a random sample $\mathcal{S}(t)$ of agents is selected from the population and each is kept in the population 
 with probability given by $p_i(t) = \frac{\pi_i(t)}{\sum_{j \in \mathcal{S}(t)}\pi_j(t)}$.
 We set $|\mathcal{S}(t)| = 10$ in this implementation.
 The mixed selection mechanism interpolates between $\mech{quantile}$ and $\mech{fps}$. 
 When a selection event occurs, with probability $\frac{1}{2}$ the mechanism $\mech{quantile}$ is used 
 and with probability $\frac{1}{2}$, $\mech{fps}$ is used.
 \medskip

 \noindent
When agents are selected out of the population, new agents are added to replace the ones that have exited 
so that the number of agents in the population is conserved. 
We set the number of agents $N_{\text{agents}} = 100$ in each run of the simulation.
When new agents enter the model after a selection event,
with probability $p_{\text{innovation}}$ 
they draw their governing parameters ($p_{\text{bid}}$, $N_{\text{shares}}$, and $\nu$)
from stationary probability distributions that do not change with selective pressure,
and with probability $1 - p_{\text{innovation}}$ they draw their governing parameters 
from the distributions of these parameters among the members 
of the population of agents that did not get selected out of the market.
In this work, we set $p_{\text{innovation}} = 0.01$.
We choose these selection mechanisms not because they are in some way optimal methods for selecting individuals
in an evolving system---in fact, the disadvantages of fitness-proportionate selection are 
well-documented \cite{whitley1989genitor}---but for their interpretation 
in the context of a financial market.
The quantile-based method models an environment in which an investing public (individuals, 
firms, etc.) actively avoid firms that are performing badly in the market, but do not actively 
seek out firms whose profits are the highest. 
In contrast, a fitness-proportionate scheme 
models a scenario in which
investors seek out the firms 
that have the highest total profits and allocate their funds to these firms in proportion to their past 
performance.
We also included a control simulation model in which no selection was present and all agents initially in the 
simulation at time $t = 0$ remained in the simulation for the entire time.

\subsection{Theoretical models}
We turn briefly to a theoretical model of the evolution of agents' parameters: 
$p_{\text{bid}}$, $N_{\text{shares}}$, and $\nu$.
For the sake of convenience we pass to a continuous time description, though the 
discrete time of the simulation is recovered by simply setting $dt = \frac{1}{24}\ \text{days}$.
We assume that prices evolve according to a zero-mean L\'{e}vy flight,
\begin{equation}\label{eq:asset-eq}
	dX(t) = \sigma_X dL^{(\alpha)}_X(t),\ X(0) = X_0,
\end{equation}
with tail exponent $\alpha \in (1.7, 2)$ as suggested by Mandelbrot \cite{mandelbrot1997variation}.
This model has been shown to give superior fit to real data when compared with the geometric Brownian 
motion model of asset prices \cite{mantegna1997econophysics,cont1997scaling}.
Since any agent whose bid probability deviates too far from the natural equilibrium of $p_{\text{bid}}^*
= \frac{1}{2}$ 
will soon become rapidly unprofitable and hence be selected out of the market, we assume $p_{\text{bid}}$ 
evolves according to a type of Ornstein-Uhlenbeck process,
\begin{equation}\label{eq:pbid-eq}
	dp_{\text{bid}}(t) = \theta_{p_{\text{bid}}}(p_{\text{bid}}^* - p_{\text{bid}}(t))\ dt
	+ \sigma_{p_{\text{bid}}}dL^{(\alpha)}_{p_{\text{bid}}}(t).
\end{equation}
In contrast, there is no logical steady state for $N_{\text{shares}}$, so we assume that its 
evolution is governed by a standard random walk with heavy-tailed increments arising from the auction 
mechanism,
\begin{equation}\label{eq:nshares-eq}
	dN_{\text{shares}}(t) = \mu_{N_{\text{shares}}}\ dt + \sigma_{N_{\text{shares}}}\ 
	dL_{N_{\text{shares}}}^{(\alpha)}(t).
\end{equation}
The parameter $\mu_{N_{\text{shares}}}$ is interpreted as evolutionary drift. 
The interpretation of volatility preference $\nu$ as a measure of risk aversion (small $\nu$) or risk 
neutrality / seeking (large $\nu$) gives insight into a possible model for its evolution.
Simply put, volatility preference increments in proportion to the current level of volatility preference:
if the population is risk averse, the variation in volatility preference should be low; 
if the population is risk neutral or risk-seeking, the variation in volatility preference will likely be
high.
Incorporating an evolutionary drift term, a reasonable model for this phenomenon is
\begin{equation}\label{eq:volpref-eq}
	d \nu(t) = \nu(t) [\mu_{\nu}  dt + \sigma_{\nu} dL_{\nu}^{(\alpha)}(t)].
\end{equation}
For example, orders submitted according to Eq.\ \ref{eq:agent-order}
with $\nu$ much larger than the population
average are unlikely to be executed if the resultant price is favorable to the submitting agent
(i.e., very high ask price or very low bid price relative to the last equilibrium price) and will result
in a large financial loss to the agent if the resultant price is likely to be executed 
(i.e., very high bid price or very low ask price).
%

\subsection{Methodology}
We seek an understanding of the effects of the selection mechanism on micro- and 
macro-market statistics. 
\begin{figure}
\centering
	\includegraphics[width=\columnwidth]{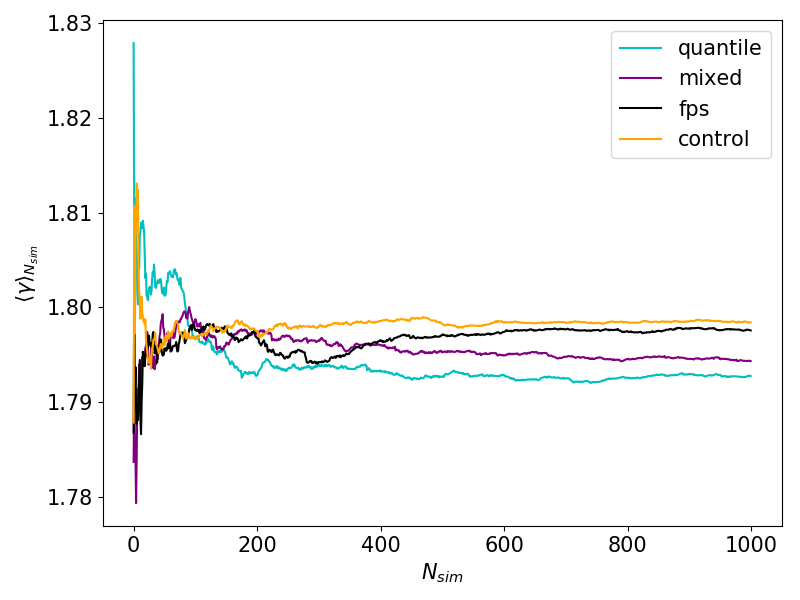}
	\caption{The mean power spectral density (PSD) exponent of population price time series,
	$\langle \gamma \rangle_{N_{\text{sim}}}= \frac{1}{N_{\text{sim}}}
	\sum_{n=1}^{N_{\text{sim}}} \gamma_n$,
	where $\gamma_n$ is defined by $S_{xx}(\omega) \sim \omega^{-\gamma_n}$.
	All PSD exponents converge to a value near 
	$\langle \gamma \rangle_{N_{\text{sim}}} \sim 1.8$, though 
	the quantile mechanism has the largest exponent and hence the 
	average price time series associated with the quantile mechanism 
	is less autocorrelated
	than the others.}
	\label{fig:psd-cgs}
\end{figure}
Are there cross-mechanism differences between optimal parameter combinations, or, more
fundamentally, is there a steady-state optimal parameter combination at all?
How do the time series of parameters---which, in a real financial market, would be unobservable---affect
macro-observable quantities such as leptokurticity of returns or volatility?
To answer these questions, we first characterize basic macro properties of the simulations under each 
selection mechanism. 
Aside from the price $X(t)$ and return $r(t) = \log_{10} X(t) - \log_{10} X(t - 1)$ time series, we
calculate the 
price power spectral density, defined by $S_{xx}(\omega) = \hat{X}(\omega)\hat{X}^{\dagger}(\omega)$,
where we have defined the Fourier transform on the interval $[0,T]$ by 
\begin{equation}
\hat{X}(\omega) = \frac{1}{\sqrt{T}}\sum_{t=1}^T X(t)e^{-i \omega t}\Delta t, 
\end{equation}
where $\Delta t = \frac{1}{24}$, so that the units of the Fourier transform are $1/\text{days}$.
For financial price time series we expect $S_{xx}(\omega) \sim \omega^{-\gamma}$, where
$\gamma \in (1.7, 2)$. 
\begin{figure}[!htp]
\centering
	\includegraphics[width=\columnwidth]{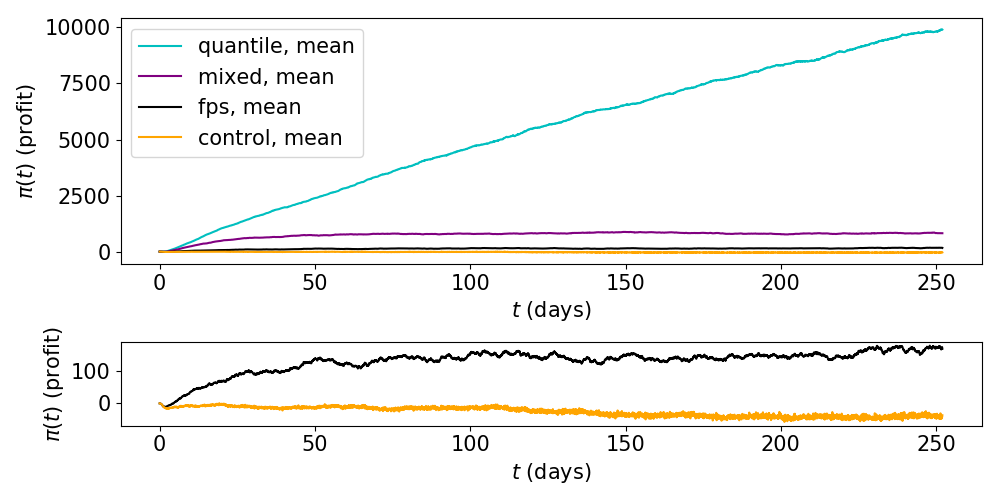}
	\caption{Mean profit levels differed by selection mechanism. 
	The quantile (truncation) selection mechanism lead to average profits that were 
	approximately an order of magnitude higher than that of the second-most profitable 
	mechanism, the mixture of fitness-proportionate selection and quantile selection. 
	While returning positive average profits, fitness-proportionate selection was the least 
	profitable of the non-control selection mechanisms.
	In this context, average profit is defined by $\langle \pi(t) \rangle_{j, \text{sim}} 
	= \frac{1}{N_{\text{sim}}N_{\text{agents}}}\sum_{n=1}^{N_{\text{sim}}}
	\sum_{j \text{ active at time $t$}}\pi_{j, n}(t)$}
	\label{fig:profit-means}
\end{figure}
Brownian motion has $\gamma = 2$, while real asset markets exhibit $\gamma \sim 1.8$ 
in price dynamics
\cite{mandelbrot1997variation,carbone2004time}.
Time series of the parameters $\pbid{j}$, $\nshares{j}$, and $\volpref{j}$ are described and 
their distributions are fit and compared with distributions predicted from the theoretical models
 described above.
Finally, we analyze the link between the agent-level
micro-volatility parameters $\volpref{j}$ and macro-volatility as measured 
from price or return time series and remark on its differentiation by selection mechanism.

\section{Results}
We ran 1000 runs of the artificial asset market simulation
for each selection mechanism (control, $\mech{quantile}$,
$\mech{fps}$, and $\mech{mixed}$) for a total of 4000 simulations.
Each simulation was composed of 24 ``hour" trading periods in each trading ``day". A total of 252 trading 
days per year (in analogy with the calendar of the U.S. national market system) resulted in a 
total of 6048 trading periods per simulation. 
The number of agents in each simulation was held constant at 100.
To determine that the number of runs of the simulation
was adequate for the calculation of population averages,
we generated reruns of the simulation until temporal averages of the population 
price time series power spectral density exponents appeared to converge.
This convergence is displayed in Figure \ref{fig:psd-cgs}.
\subsection{Profitability and parameter evolution}
The mean profitability of agents under each selection mechanism is displayed in Figure 
\ref{fig:profit-means}.
Here, we define an average over both runs of the simulation and active agents, {\it viz.}
\begin{equation}
	\langle \pi(t) \rangle_{j,\ \text{sim}} = 
	\frac{1}{N_{\text{agents}}N_{\text{sim}}}\sum_{n=1}^{N_{\text{sim}}}
	\sum_{j \text{ active at time $t$}}\pi_{j,n}(t).
\end{equation}
The purely quantile-based mechanism displays average profitability that is over an order of magnitude 
greater than either $\mech{mixed}$ or $\mech{fps}$,
while $\mech{mixed}$ was still much more profitable on average than was $\mech{fps}$.
This differentiation is likely to due to the fact that $\mech{quantile}$ selects out the ten worst-performing
individuals each time it is active, while $\mech{fps}$ selects out on average $|\mathcal{S}(t)| - \sum_{j 
\in \mathcal{S}(t)} p_j = 9$ individuals that are randomly sampled from the population; while the individuals 
selected out are, on average, the worst performing individuals in that particular $\mathcal{S}(t)$, they 
are by no means the worst-performing individuals in the entire population.
Though this implementation of $\mech{fps}$ results in
significantly less selective pressure on the population 
than does $\mech{quantile}$,
this choice is made to hold constant the number of individual 
agents involved in the selection step of the 
market simulation.
\medskip

\noindent
Agents' parameters---the probability of submitting a bid, $\pbid{j}$,
the mean number of shares submitted in an order $\nshares{j}$, and the
volatility preference $\volpref{j}$---were influenced by the choice of selection mechanism.
Overall, $\mech{quantile}$ was associated with lower standard deviations of parameter time series 
as calculated over runs of the simulation. 
Figure \ref{fig:params-tile} displays parameter time series for all runs of the simulation in the left panel,
and averages over runs of the simulation in the right panel.
Both $\mech{quantile}$ and $\mech{mixed}$ showed time decay toward lower values 
in $N_{\text{shares}}$ and 
$\nu$ when averaged over both active agents and runs of the simulation.
On the contrary, $\mech{fps}$ showed no decay in either parameter when the same average was performed. 
When decoupled from time, distributions of the parameters showed remarkable similarity across mechanisms, 
showing evidence for a unified underlying evolutionary model as proposed in Eqs.\ \ref{eq:pbid-eq} - 
\ref{eq:volpref-eq}, the parameters of which depend on the selection mechanism.
These time-decoupled distributions are displayed in Figure \ref{fig:params-dists}.
\subsection{Volatility correlation}
Since it seems reasonable that a fitness-proportionate selection mechanism most closely approximates
the selection mechanism operating in today's financial asset markets, 
we are particularly interested in correlations between micro-volatility---agents' volatility preferences 
$\volpref{j}$---and macro measures of volatility.
We are interested in the effects of mechanism on these macro measures of volatility, and particularly 
wish to test if micro-volatility is correlated with macro-volatility in the cases of $\mech{fps}$
and $\mech{mixed}$,
as this could provide some insight into how volatility is generated in real financial markets.
\begin{figure}
\centering
	\includegraphics[width=\columnwidth]{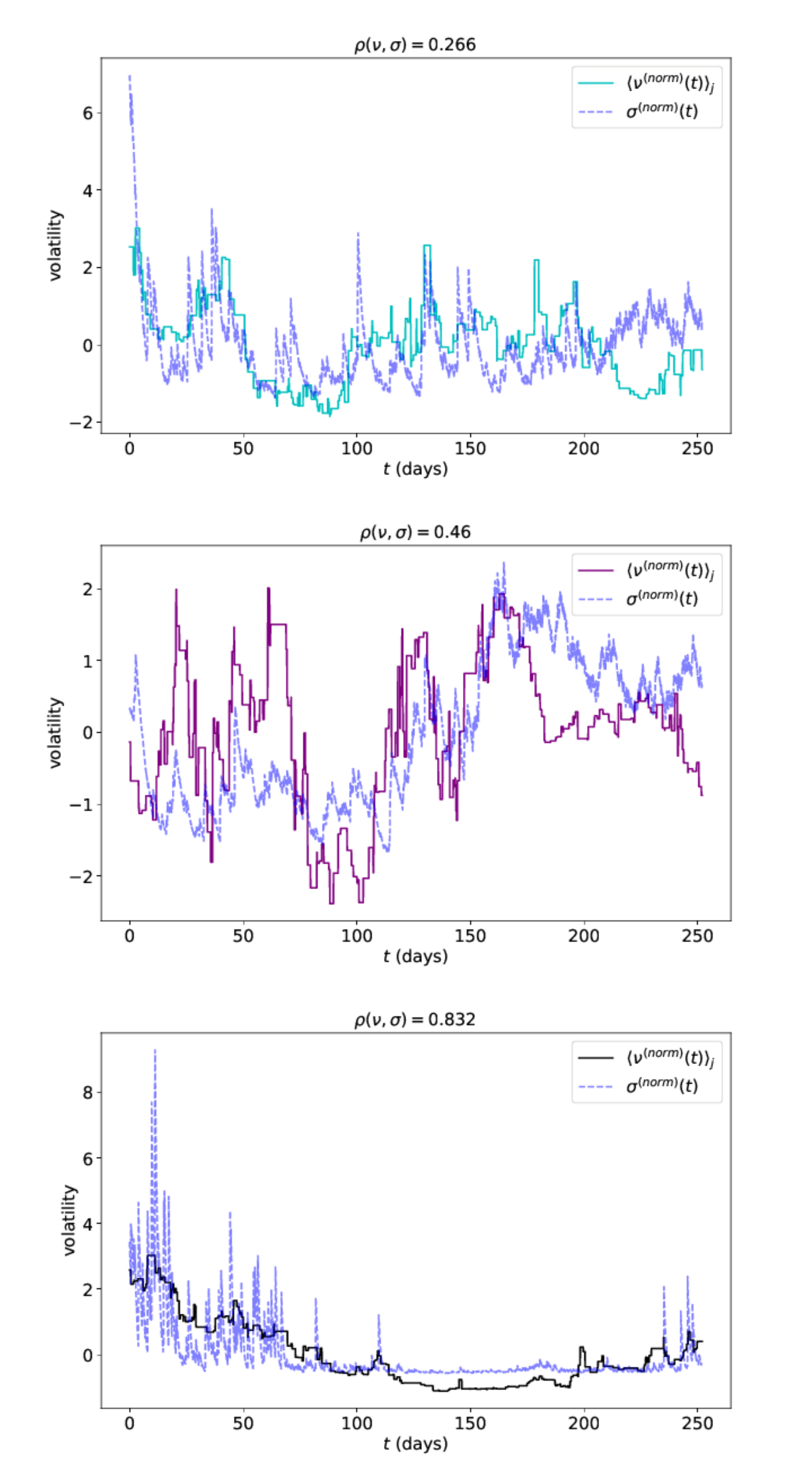}
	\caption{Micro-macro volatility correlation varies by selection mechanism. 
	We chose an arbitrary rerun and show the average volatility preference,
	$\langle \nu(t) \rangle_j = \frac{1}{N_{\text{agents}}} \sum_{j \text{ active at time $t$}}
	\nu_j(t)$, displayed as a solid curve, 
	plotted against macro-volatility calculated as the solution of a $\text{GARCH}(1,1)$ process,
	displayed as a dashed curve.
	After calculation, these processes were normalized to have zero mean and unit variance 
	for display on the same scale.
	From top to bottom: $\mech{quantile}$, 
	$\mech{mixed}$, and $\mech{fps}$.}  
	\label{fig:vol-montage}
\end{figure}
\begin{figure}
\centering
	\includegraphics[width=\columnwidth]{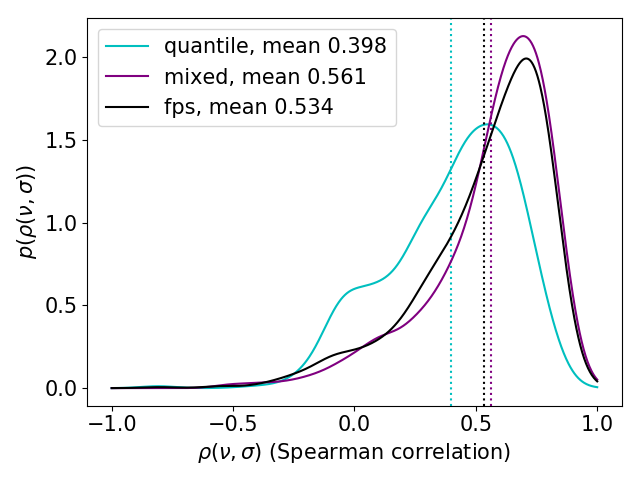}
	\caption{Micro and macro volatility measures are highly correlated when fitness-proportionate
	selection is included in the selection mechanism (i.e., the mechanism is either 
	mixed or fitness-proportionate).
	There is correlation between micro and macro volatility under the pure quantile mechanism, 
	but the effects of agents' volatility preferences are muted in comparison.
	Calculated values were used in a kernel density estimate, plotted above, 
	computed using Gaussian kernels and the Silverman rule for bandwidth estimation.}
	\label{fig:micro-macro-vol-pdf}
\end{figure}
Macro-volatility---volatility as measured from market-wide statistics such as price and returns---is 
often modeled using a generalized autoregressive conditional heteroskedasticity (GARCH) model 
\cite{bollerslev1986generalized}, 
which, in its most basic form,
hypothesizes that log returns $r(t) = \log_{10}X(t) - \log_{10}X(t - 1)$ can be decomposed as 
\begin{align}
	r(t) &= \mu + \varepsilon(t)\\
	\varepsilon(t) &= \sigma(t) z(t)\\
	\sigma^2(t) &= \xi + \alpha \varepsilon^2(t-1) + \beta \sigma^2(t-1),
\end{align}
where $z(t) \sim \mathcal{N}(0,1)$.
For each simulation, we compute a GARCH model of the form given above and calculate the Spearman 
correlation coefficient $\rho(\langle \nu \rangle, \sigma)$ between the average agent volatility preference 
$\langle \nu(t) \rangle_j = \frac{1}{N_{\text{agents}}}\sum_{j \text{ active at time $t$}} \nu_j(t)$ 
and the fitted volatility $\sigma(t)$.
Figure \ref{fig:vol-montage} displays $\langle \nu(t) \rangle_j$ and $\sigma(t)$ for an arbitrarily 
chosen run of the simulation. 
Figure \ref{fig:micro-macro-vol-pdf} displays the empirical probability density function (pdf)
of $\rho(\langle \nu \rangle_j, \sigma)$ across all non-control simulations. (The pdf of correlations 
for the control is sharply peaked about zero and uninteresting as 
there is no evolution of $\volpref{j}$ in this case.)
The pdf of correlation coefficients for $\mech{quantile}$ is bimodal, with one mode about zero 
and another near $\rho = 0.5$,
while for $\mech{mixed}$ and $\mech{fps}$ the pdfs are are peaked near $\rho \simeq 0.75$ with a long 
left tail.
\subsection{Theoretical fit}
Since the theoretical models for the evolution of agents' parameters given by 
Eqs.\ \ref{eq:pbid-eq} - \ref{eq:volpref-eq} contain nine free parameters in total, 
to assess their suitability as a first-order theoretical model of the evolutionary 
phenomena occurring here we must fit these parameters from the data generated by the agent-based 
model.
\begin{figure}[!htp]
\centering
	\includegraphics[width=\columnwidth]{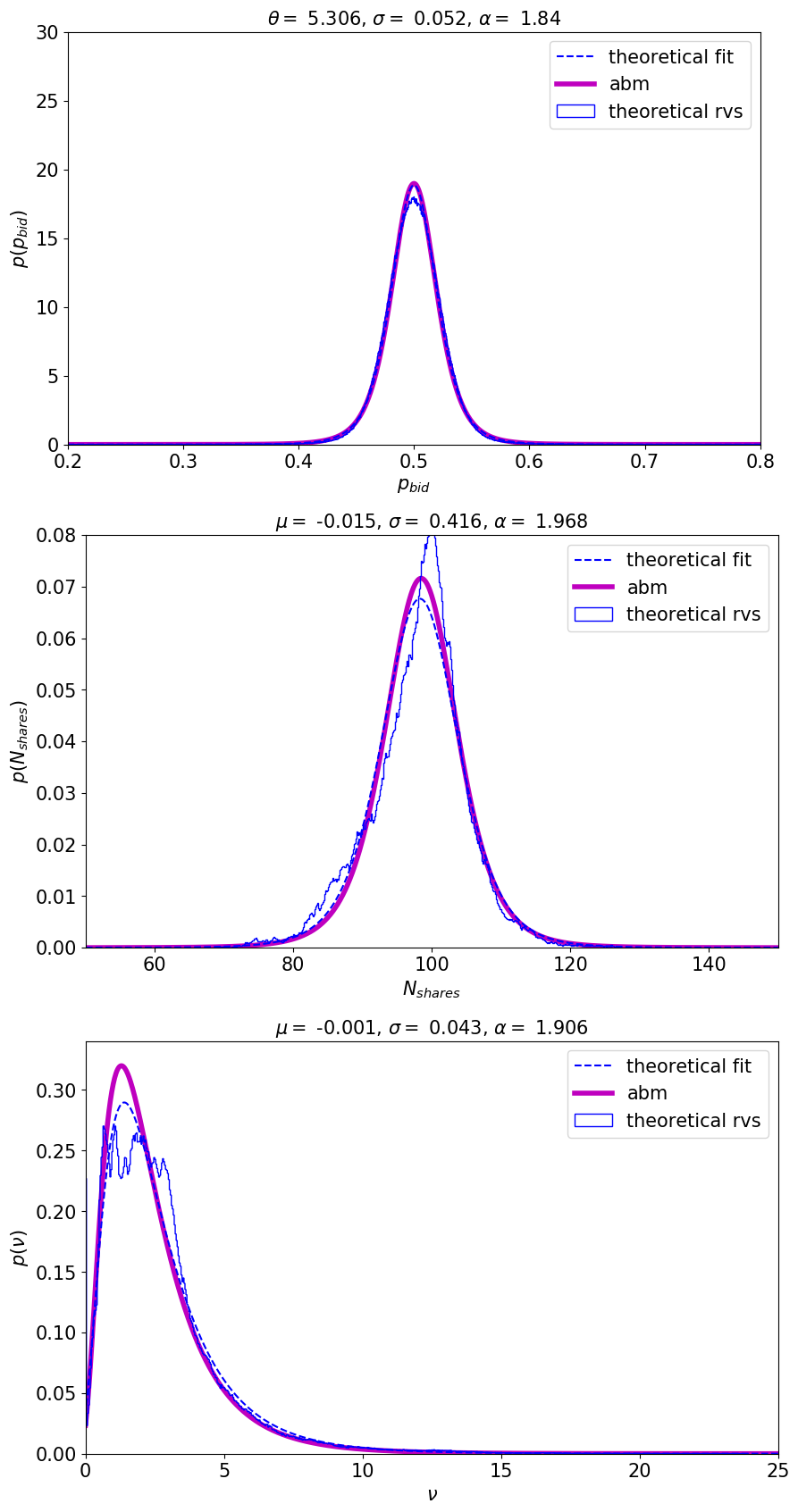}
	\caption{Parameters to theoretical models of $p_{\text{bid}}$,
	$N_{\text{shares}}$,
	and $\nu$ were fit using maximum likelihood estimation and differential evolution,
	as described in the text.
	Displayed here are the fit distributions of the theoretical models for the mixed 
	mechanism in dashed blue curves, 
	random variates drawn from the theoretical model in solid blue curves,
	and fit distributions of the ABM in magenta curves.,
	Calculated optimal values of free parameters for each model 
	are displayed in the title of each panel.}
	\label{fig:theo-params}
\end{figure}
To do this we hypothesize a parametric form $p_{\text{theo}}(x|\beta)$ 
for each distribution: $p(p_{\text{bid}})$, $p(N_{\text{shares}})$, and 
$p(\nu)$.
The optimal values of $\beta$ are defined as the vector that minimizes
\begin{equation}\label{eq:dkl}
	\int\displaylimits_{x \in \Omega} p_{\text{abm}}(x)\log \left(\frac{p_{\text{abm}}(x)}
	{p_{\text{theo}}(x|\beta)}\right)\ dx,
\end{equation}
the Kullback-Leibler (KL) divergence of the theoretical distribution away from the distribution
produced by the ABM.
The domain of integration $\Omega$ is defined as 
all observed values of the quantity $x$ for each time step 
and each run of the simulation. 
This integral is minimized using differential evolution \cite{storn1997differential},
at each iteration of which a number of simulations of the theoretical model Eqs.\ \ref{eq:pbid-eq} - 
\ref{eq:volpref-eq} are calculated and the maximum likelihood estimation of the parameter vector 
$\beta$ is found, which is then substituted into the functional form of $p_{\text{theo}}$ used in the 
definition of KL divergence.
\medskip

\noindent
Figure \ref{fig:theo-params} displays comparisons between the fitted theoretical distributions 
and distributions arising from the ABM for $\mech{mixed}$.
To emphasize that the restriction of the fit distribution to a parameterized form does not result in a 
model that fits the data poorly, 
random variates drawn from each model are drawn and their histogram is plotted along with the 
fit distributions.
The calculated optimal values of the free parameters for each model are displayed in the title of each 
panel.
There is strong restorative force ($\theta_{p_{\text{bid}}} = 5.306$) to the 
equilibrium bid probability $p_{\text{bid}} = \frac{1}{2}$,
while there is negative evolutionary drift in mean number of shares submitted per order 
($\mu_{N_{\text{shares}}} = -0.015$) and volatility preference ($\mu_{\nu} = -0.001$).

\section{Discussion and conclusion}
We find that choice of selection mechanism is associated with differential behavior of asset price 
spectra, agent parameter distributions and time series, and volatility. 
While the probability of submitting a bid order fluctuates regularly about its natural equilibrium value 
of $p_{\text{bid}}^* = \frac{1}{2}$ under all three mechanisms, 
the time series of the average number of shares traded and the volatility preference parameter varies 
functionally depending on the presence of a quantile-based component to the selection mechanism. 
When a quantile-based component is not present ($\mech{fps}$), these time series vary
in the mean case very little from 
their initial values, with a slight upward trend.
However, when a quantile-based component is present, in the mean case these series exhibit a steady trend
toward lower values. 
In both $N_{\text{shares}}$ and $\nu$, $\mech{mixed}$ trends most strongly toward lower values and does not 
appear to converge in the time period covered by our simulation (252 days of trading once per hour), 
suggesting that longer simulation run times are necessary to discern the nature of the steady state of these 
parameters under mechanisms containing a quantile-based component, if such steady-states exist.
\medskip

\noindent
All three mechanisms show significant correlation between micro-volatility, as measured by 
the risk-aversion / volatility preference parameter $\nu$,
and market-wide volatility measured from 
the market price using standard econometric models (GARCH).
All distributions of Pearson correlation coefficients of micro- and macro-volatility exhibited
negative skew 
(more weight in the left-hand tail).
The quantile-based mechanism displayed bimodality in this distribution, with a small peak near zero 
correlation and a large peak near $\rho = 0.5$.
Contrasting with this, $\mech{mixed}$ and $\mech{fps}$ were unimodal, with peaks near $\rho \simeq 0.75$, 
displaying a strong median correlation between micro- and macro-volatility.
\medskip

\noindent
Taken together, these results paint a picture of nontrivial interaction between selection mechanism and 
market outcomes. Mechanisms that include a fitness-proportionate component show higher volatility than 
a purely quantile-based mechanism, and under those mechanisms micro-volatility is more highly correlated 
with observable macro-volatility, providing a possible mechanistic explanation for the generation of 
macro-volatility in real financial markets.
However, mechanisms that contain a quantile-based component show significant evolutionary drift 
in the average number of shares submitted per order and in volatility preference. 
When taken along with the fact that these mechanisms produced far higher average profits than did the purely 
fitness-proportionate method, this suggests that lower values of these parameters are---in a population 
of zero-intelligence agents, at least---associated with higher average profit levels, possibly 
due to an increase in risk-aversion among the population of agents and a corresponding decrease in the 
frequency of agents that experience massive trading losses.
\medskip

\noindent
Our study has several areas on which future work could improve, the most important of which being
our neglection of other selection mechanisms. 
There are far more---and more realistic!---mechanisms that provide a model for how agents may be removed 
from, and added to, a financial market. Drawing definitive conclusions about the nature of market selection 
and competition from a study of only two fundamental mechanisms
is ill-advised, and we decline to do this.
Another shortcoming is our lack of variation of many parameters in this study. 
In order to understand these mechanisms in more depth, a detailed study of macro-observable market 
statistics as a function of, e.g., tournament size, quantile, and mixture probability between the 
two fundamental mechanisms is required.
Future work should focus on inclusion of more and different selection mechanisms, as well as inclusion 
of more advanced agents.

\begin{acks}
The authors are grateful for helpful conversations with
Laurent H\'{e}bert-Dufresne,
Tyler John Gray, Brian F.\ Tivnan, Peter Sheridan Dodds, 
Chris Danforth, John Henry Ring IV, Sage Hahn, and Josh Bongard,
and thankful for the insightful comments provided by three anonymous referees.
\end{acks}

\bibliographystyle{ACM-Reference-Format}
\bibliography{sample-bibliography}

\end{document}